# Determination of Burgers-vector directions of threading edge dislocations in GaN by phase-contrast microscopy

Yukari Ishikawa[1,*], Maki Shimizu[1], Koji Sato[1], Kisara Matsumoto[2], Kazuki Ohnishi[2], Yongzhao Yao[1,2], Ryo Hattori[3]

[1]Japan Fine Ceramics Center, 2-4-1, Mutsuno, Atsuta, Nagoya, 456-8587, Japan

[2]Mie University, 1577, Kurimamachiya-cho, Tsu, Mie 514-8507, Japan

[3]Ceramic Forum Co., Ltd, #502, 2-4-14 Sakaemachidori, Chuo, Kobe 650-0023, Japan

Phase-contrast microscopy enables rapid, nondestructive observation of threading dislocations in GaN, but the relationship between dislocation contrast and the Burgers vector has not been established. Here, we demonstrate that the in-plane Burgers-vector direction of threading edge dislocations in GaN can be determined from the arrangement of bright and dark contrast in phase-contrast microscopy images. The relationship was first established for threading edge dislocations in ammonothermal GaN whose Burgers vectors were independently determined by X-ray topography and was subsequently applied to hydride vapor phase epitaxy-grown GaN. The Burgers vector predicted from the phase contrast was verified by weak-beam dark-field TEM and large-angle convergent-beam electron diffraction. For the threading edge dislocations examined, the direction obtained by rotating the direction from the minimum of the dark contrast to the maximum of the bright contrast by 90° counterclockwise coincided with the Burgers-vector direction. Analysis based on the photoelastic effect further suggests that the phase contrast primarily reflects the sum of the two in-plane normal strain components, $\varepsilon_{xx} + \varepsilon_{yy}$, around the dislocation. These results demonstrate that phase-contrast microscopy can be used as a nondestructive optical method for determining the in-plane Burgers-vector direction of threading edge dislocations in GaN.

*Corresponding author

GaN, one of the wide-bandgap semiconductors, possesses excellent properties such as a high breakdown field, high carrier mobility, and a direct bandgap, and is therefore expected to be used not only in light sources such as blue LEDs and lasers but also in next-generation power devices[1–4]. However, GaN crystals still contain many defects, which can degrade device performance, lifetime, and reliability[5–9]. A technique capable of nondestructively inspecting an entire wafer within a short time is important not only for providing essential knowledge for the growth of low-defect-density crystals but also for device manufacturing.

For nondestructive and rapid inspection of an entire wafer, optical-microscopy-based techniques that can acquire images over a wide area in a single shot are particularly suitable. Threading dislocations in wide-bandgap semiconductors have been detected using polarized-light microscopy[10–20] and phase-contrast microscopy[21–24]. In polarized-light microscopy, the dislocation contrast has been reported to resemble the shear-strain distribution around a dislocation, and the in-plane direction of the Burgers vector can be determined from the arrangement of bright and dark contrast[11–19].

Although dislocations have been observed by phase-contrast microscopy, the relationship between the contrast and the Burgers vector has not yet been clarified, and no method for dislocation classification has been reported[21–24]. Phase-contrast microscopy, however, provides higher spatial resolution than polarized-light microscopy, enabling individual dislocations to be distinguished even in substrates with high dislocation densities ($3\times10^{6}$ $cm^{-2}$). It can also simultaneously visualize multiple types of defects, including scratches, voids, and facet domains, and is therefore promising for comprehensive whole-wafer defect inspection[24]. Establishing a method for determining the Burgers vector of dislocations by phase-contrast microscopy is thus important for realizing comprehensive wafer evaluation using a single inspection technique.

In this study, we verified by X-ray topography (XRT) and TEM-large-angle convergent-beam electron diffraction (LACBED) analysis that the in-plane component of the Burgers vector of threading edge dislocations in GaN can be determined from the arrangement of bright and dark contrast in phase-contrast microscopy images. Furthermore, to investigate the contrast-formation mechanism in phase-contrast microscopy, the experimentally observed contrast distribution around dislocations was compared with a theoretical analysis of refractive-index changes. The results suggest that the phase contrast reflects changes in the average refractive index caused by the strain field of the dislocation.

Two types of samples were used: a low-dislocation-density ammonothermal GaN wafer, for which the Burgers vector could be determined by XRT, and a high-dislocation-density hydride vapor phase epitaxy (HVPE)-grown GaN wafer. The two samples exhibited different dislocation structures and were therefore used for different purposes. The commercially available 2-inch GaN (0001) wafer grown by a conventional ammonothermal method[25] was used. The Ga face had been chemical mechanical polish (CMP)-finished, and the N face was CMP-finished after receipt of the sample. The wafer thickness was approximately 340 μm, and the [0001] axis was tilted by 0.5° from the wafer normal toward [$\bar{1}$100] by rotation about the [11$\bar{2}$0] axis. The threading-dislocation density was approximately $3 \times 10^4$ $cm^{-2}$, allowing the Burgers vector of individual dislocations to be determined by XRT. This is the same sample analyzed in Ref. 20. A chip was cut from a commercially available 2-inch HVPE-grown GaN (0001) wafer; it was taken from the same wafer as the chip analyzed in Ref. 24. Because the dislocation density was $7 \times 10^5$ $cm^{-2}$, discrimination of individual dislocations by XRT was difficult.

Phase-contrast microscopy was performed using a Crystalline Tester CP1 (Ceramic Forum Co., Ltd., Japan) equipped with a 20× CFI Plan Fluor DLL positive-phase-contrast objective lens (NA = 0.5, PH1; Nikon, Japan), a ring-aperture condenser (NA = 0.78), and a 405-nm LED light source. Only brightness and contrast adjustments were applied to the images; no other image processing was performed.

Reflection XRT images of the ammonothermal GaN wafer were acquired using six equivalent $\boldsymbol{g}$ = 11−24 reflections at BL3C of the High Energy Accelerator Research Organization (KEK). The X-ray wavelength was $\lambda$ = 1.4 Å, with an incidence angle $\alpha$ = 5° and a diffraction angle $\theta$ = 44°. For the HVPE-grown GaN chip, etch pits were formed by immersion in a KOH + NaOH melt at 450 °C, and a TEM specimen containing the target etch pit was prepared by FIB microsampling. The dislocation was characterized by weak-beam dark-field imaging and large-angle convergent-beam electron diffraction (LACBED) analysis to determine its Burgers vector[26–30].

Figure 1(a) shows a phase-contrast microscopy image around a honeycomb defect previously reported in ammonothermal GaN[25,31]. The arrangement of bright and dark dislocation contrast did not change as the focal position varied from the front surface to the back surface. Because the contrast was clearer inside the crystal than at the surface, the image shown here was acquired with the focus positioned approximately 48 μm below the surface.

The honeycomb defect observed here is the same defect previously examined by polarized-light microscopy in Ref. 20. Honeycomb defects have been reported to consist of threading-edge-dislocation bundles arranged in a hexagonal configuration[31]. Therefore, the Burgers-vector direction can be uniquely determined from the bright/dark contrast in XRT images acquired using six $\boldsymbol{g}$ vectors equivalent to $11\bar{2}4$. The Burgers vectors of the individual dislocation bundles were determined using the convention shown in Supplementary Fig. 1 and are indicated by arrows $\boldsymbol{b}_1$–$\boldsymbol{b}_6$ in Fig. 1(a) (see Table I of Ref. 20 for details).

In addition, isolated dislocations whose XRT bright/dark contrast and trace directions matched those of one of $\boldsymbol{b}_1$–$\boldsymbol{b}_6$ were assigned the same in-plane Burgers vector as the corresponding dislocation bundle[20].

Figure 1(b) shows, from left to right for dislocations corresponding to each Burgers vector, an enlarged phase-contrast image of a dislocation bundle (DB), an enlarged image of an isolated dislocation, an image of the isolated dislocation (ID) indicating the position of the extracted contrast profile (yellow line), and the corresponding contrast profile. The relationship between phase contrast and the Burgers vector is described using $\boldsymbol{b}_1$ as an example. For $\boldsymbol{b}_1$, both the dislocation bundle and the isolated dislocation exhibited dipole-like contrast that was dark at the lower left and bright at the upper right, with the boundary between the bright and dark regions parallel to the Burgers-vector direction. The contrast profile acquired along the yellow line from the white triangle toward the black triangle showed a clear dark-to-bright reversal. The direction obtained by rotating the direction from the minimum of the dark contrast to the maximum of the bright contrast by 90° counterclockwise coincided with the Burgers-vector direction.

The same relationship was also observed for $\boldsymbol{b}_2$ and $\boldsymbol{b}_3$. For $\boldsymbol{b}_4$, $\boldsymbol{b}_5$, and $\boldsymbol{b}_6$, linear contrast extending toward the lower left, downward, or lower right, respectively, was superimposed on the dipole-like contrast. The Burgers vector could still be determined for $\boldsymbol{b}_5$, whereas for $\boldsymbol{b}_4$ and $\boldsymbol{b}_6$, the superimposed linear contrast made it difficult to determine the direction of the dipole-like contrast.

As reported in Ref. 20, the **c** axis is tilted by 1.3° from the optical axis toward $[\bar{1}100]$, while the TEDs are tilted by 3.3° from the c axis in directions perpendicular to their Burgers vectors. The resulting dislocation-line inclinations relative to the optical axis are 2.1° for $\boldsymbol{b}_2$, 2.9° for $\boldsymbol{b}_1$ and $\boldsymbol{b}_3$, 4.3° for $\boldsymbol{b}_4$ and $\boldsymbol{b}_6$, and 4.5° for $\boldsymbol{b}_5$. Phase contrast has been reported to change from point-like to linear with increasing dislocation-line inclination[24]; thus, the larger inclinations for $\boldsymbol{b}_4$–$\boldsymbol{b}_6$ account for the superimposed linear contrast.

Because many TEDs in the ammonothermal sample used here are inclined by approximately 3° in a direction perpendicular to their Burgers vectors (Supplementary Fig. 2(a))[20], the dislocation-line inclination is correlated with the Burgers vector. Therefore, this sample alone cannot distinguish whether the phase-contrast pattern is governed by the Burgers vector or by the inclination of the dislocation line. We therefore performed the same analysis on an HVPE-grown GaN sample in which no correlation was observed between the dislocation-line propagation direction and the Burgers vector (Supplementary Fig. 2(b)).

Figure 2(a) shows a phase-contrast microscopy image of the HVPE-grown crystal. In this crystal, threading edge dislocations are generally nearly parallel to the optical axis and therefore exhibit point-like contrast, whereas many threading mixed dislocations are inclined by approximately 8° to the optical axis and exhibit linear contrast[24]. To examine the relationship with the Burgers vector, the following analysis focused on TEDs showing point-like contrast. The bright/dark arrangement of the point-like contrast was confirmed to remain unchanged from the front surface to the back surface. Because the arrangement was clearer inside the crystal than at the surface, the image shown was acquired with the focus approximately 36 μm below the surface. At least six types of bright/dark contrast arrangements, labeled i–vi, were identified. Representative dislocations from each group are indicated by white arrows in Fig. 2(a). Figure 2(b) shows six sets for representative point-like contrasts selected from groups i–vi: an enlarged image, an image indicating the position used to extract the contrast profile, and the corresponding cross-sectional contrast profile. Type i exhibits dipole-like contrast that is dark at the lower left and bright at the upper right. Applying the criterion obtained from Fig. 1, this contrast corresponds to Burgers vector $\boldsymbol{b}_1$. The cross-sectional profile also changes from a minimum to a maximum from left to right. Types ii–vi are similarly predicted from their bright/dark arrangements to correspond to $\boldsymbol{b}_2$–$\boldsymbol{b}_6$ in Fig. 1. In addition, iv, v, and vi show contrast polarity opposite to i, ii, and iii, respectively, suggesting mutually opposite Burgers vector directions. These results demonstrate that the Burgers vector can also be predicted from the bright/dark arrangement of phase contrast in HVPE-grown GaN. To verify this prediction, representative dislocations were analyzed using TEM techniques.

Figure 3(b) shows the region in Fig. 3(a) (reproduced from Fig. 2(a)) after etching. The resulting pits approximately coincide with the phase-contrast features. The pit corresponding to the contrast indicated by the arrow in region v of Fig. 3(a) [Fig. 3(c)] was selected and sampled by FIB for cross-sectional TEM and LACBED observation along $[1\bar{1}00]$ [Fig. 3(d)]. Figure 4(a) shows the dislocation beneath the etch pit observed under a zone-axis condition. Figures 4(b) and 4(c) show weak-beam dark-field images acquired under ***g***–3***g*** conditions using $\boldsymbol{g}$ = 0002 and $\boldsymbol{g}$ = $\bar{1}\bar{1}20$, respectively. The dislocation contrast disappears in Fig. 4(b) but remains visible in Fig. 4(c), indicating that the Burgers vector has no c component. In a LACBED pattern, the number of nodes n formed at the intersection between a diffraction line $\boldsymbol{g}$ and a Burgers vector $\boldsymbol{b}$ satisfies $\boldsymbol{g}\cdot\boldsymbol{b}$ = n.[26–30] No node is observed at the intersection of the dislocation line and the $\boldsymbol{g}_1$ = 0008 diffraction line, confirming that $\boldsymbol{b}$ contains no **c** component. In contrast, $n_2$ = 4 and $n_3$ = −5 nodes are observed for the $\boldsymbol{g}_2 = 22\bar{4}\bar{8}$ and $\boldsymbol{g}_3 = \bar{3}\bar{2}5\bar{5}$ diffraction lines, respectively. Solving these equations gives $\boldsymbol{b} = 1/3[11\bar{2}0]$, which coincides with the direction obtained by rotating the direction from dark to bright by 90° counterclockwise. Thus, the Burgers vector of a threading edge dislocation can be determined from its phase contrast. Interestingly, the Burgers vectors of the dislocation arrays labeled i–vi in Fig. 2(a) are parallel to the arrays. This is difficult to reconcile with a simple low-angle grain-boundary dislocation arrangement and instead suggests that the arrays may originate from misfit dislocations extending to the surface.·

For the threading edge dislocations observed in this study, the Burgers-vector direction coincided with the direction obtained by rotating the direction from the minimum of the dark phase contrast to the maximum of the bright contrast by 90° counterclockwise. A similar relationship has been reported in Raman mapping around dislocations by Kokubo et al.[32–35]. Specifically, rotating the direction from the minimum of the low-wavenumber shift to the maximum of the high-wavenumber shift of the Raman $E_2^H$ peak by 90° counterclockwise gives the corresponding Burgers-vector direction. This agreement suggests that phase-contrast microscopy and Raman spectroscopy may reflect a common strain component around the dislocation. We therefore examine below the strain components reflected by the two techniques. The Raman peak shift (Δω) around an edge dislocation is expressed as follows[33, 34].

$$\Delta\omega = a\left(\varepsilon_{xx} + \varepsilon_{yy}\right) + b\varepsilon_{zz} \qquad (1)$$

Here, the strain terms ($\varepsilon_{ij}$) represent the strain field generated by the edge dislocation, and $a$ and $b$ are phonon deformation-potential constants. Assuming a straight edge dislocation with its line nearly parallel to the $c$-axis and neglecting surface relaxation, $\varepsilon_{zz}$can be approximated as zero[36, 37]. Therefore, the spatial distribution of the Raman peak shift primarily reflects that of the in-plane strain, $\varepsilon_{xx} + \varepsilon_{yy}$.

The phase change detected by phase-contrast microscopy is given by

$$\phi(x, y) = 2\pi/\lambda \cdot \int \Delta n_{eff}\,(x, y, z)dz \qquad (2)$$

Thus, if the effective information-acquisition depth of the objective lens is denoted by $l$,

$$\phi(x, y) = 2\pi/\lambda \cdot \Delta n_{eff} l \qquad (3)$$

Here, $l$ represents the effective axial information-acquisition depth of the objective lens, rather than the physical propagation distance from the specimen surface to the focal plane. In our previous phase-contrast microscopy study using a comparable optical configuration, this depth was estimated to be approximately 43 μm.[24] Thus, changing the focal position primarily shifts the axial region contributing to the image rather than increasing the effective integration length. Equation (3) therefore describes the phase shift accumulated over this finite axial response range.

The in-plane (XY) part of the optical impermeability tensor is $2 \times 2$

$$\mathbf{B}_{xy} = \begin{pmatrix} B_{11} & B_{12} \\ B_{12} & B_{22} \end{pmatrix} \qquad (4)$$

The eigenvalues of this tensor are

$$\lambda_{\pm} = (B_{11} + B_{22})/2 \pm \sqrt{((B_{11} - B_{22})/2)^2 + B_{12}^2} \qquad (5)$$

and the refractive indices corresponding to the major and minor axes of the index ellipsoid are

$$n_{\pm} = 1/\sqrt{\lambda_{\pm}}. \quad (6)$$

Taking the unstrained state as the reference, the in-plane optical impermeability components are $B_{11} = B_{22} = 1/n_0^2$ and $B_{12} = 0$, where $n_0$ is the refractive index in the XY plane. For small strain-induced changes, the changes in the eigenvalues are given by,

$$\Delta\lambda_{\pm} = (\Delta B_{11} + \Delta B_{22})/2 \pm \sqrt{((\Delta B_{11} - \Delta B_{22})/2)^2 + \Delta B_{12}^2} \quad (7)$$

The corresponding changes in the principal refractive indices are, to first order,

$$\Delta n_{\pm} \simeq -n_o^3/2 \cdot \Delta\lambda_{\pm}. \quad (8)$$

In the phase-contrast microscope used in this study, no specific polarization component was selected by a polarizer or analyzer. Therefore, as a first-order approximation, we assume that the observed phase change is governed by the change in the average of the two in-plane principal refractive indices

$$\Delta \mathrm{n}_{eff} = (\Delta n_{+} + \Delta n_{-})/2 = -n_o^3/2 \cdot (\Delta B_{11} + \Delta B_{22})/2. \quad (9)$$

Here, the impermeability changes are expressed using the photoelastic tensor $p_{\mathrm{ij}}$. For a straight edge dislocation with its line nearly parallel to the $c$-axis, the out-of-plane strain component can be approximated as zero ($\varepsilon_{zz} \approx 0$). They can therefore be written as

$$\Delta B_{11} = p_{11}\varepsilon_{xx} + p_{12}\varepsilon_{yy}, \quad (10)$$

$$\Delta B_{22} = p_{12}\varepsilon_{xx} + p_{11}\varepsilon_{yy}. \quad (11)$$

Therefore, the change in the average refractive index is proportional to the sum of the in-plane strain components:

$$\Delta \mathrm{n}_{eff} = -n_0^3/4 \cdot (p_{11} + p_{12})(\varepsilon_{xx} + \varepsilon_{yy}). \quad (12)$$

This result suggests that both phase-contrast microscopy and Raman spectroscopy primarily reflect the in-plane strain distribution around a dislocation, explaining why similar relationships are obtained between the polarity of the contrast (or peak shift) and the Burgers-vector direction in the two techniques. However, because experimental values of the photoelastic tensor $p_{ij}$ of GaN have not been reported, quantitative evaluation of the sign and magnitude of the refractive-index change is currently difficult. Nevertheless, an order-of-magnitude estimate using reported theoretical photoelastic coefficients indicates that the resulting strain-induced phase modulation is much smaller than 1 rad under the present optical conditions (see Supplementary Material). In addition, changes in the absorption coefficient caused by strain-induced bandgap changes may exhibit a similar spatial distribution. In this study, the contrast-formation mechanism was discussed on the basis of the photoelastic effect; the contribution of absorption changes requires further investigation.

We demonstrated that the in-plane component of the Burgers vector of threading edge dislocations in GaN (0001) wafers can be determined by phase-contrast microscopy. In the phase contrast of a dislocation, the direction obtained by rotating the direction from the minimum of the dark contrast to the maximum of the bright contrast by 90° counterclockwise coincided with the corresponding Burgers-vector direction. Comparison with the theoretical analysis suggested that the phase contrast of the dislocations primarily reflects the distribution of the in-plane strain $(\varepsilon_{xx} + \varepsilon_{yy})$ induced by the dislocations through the photoelastic effect. When the inclination of a threading edge dislocation with respect to the optical axis becomes large, linear contrast arising from the inclined dislocation line is superimposed on the dipole-like contrast, making determination of the Burgers vector difficult.

## Supplementary Material

See the supplementary material for the definition of the Burgers-vector convention used in this study, projection images showing the dislocation-line inclinations in ammonothermal and HVPE-grown GaN, and an order-of-magnitude estimate of the strain-induced phase shift based on the photoelastic effect.

## Acknowledgments

This study was partially supported by the METI Monozukuri R&D Support Grant Program for SMEs (Grant Number JPJ005698) and the Japan Society for the Promotion of Science KAKENHI (Grant No. 23K04444, 23K17356).

## Author Declarations

## Conflict of Interest

The authors have no conflicts to disclose.

## Ethics Approval

This study did not involve human participants or animal subjects and therefore did not require ethics approval.

## Author Contributions

**Yukari Ishikawa**: Conceptualization (lead); Data curation (equal); Formal analysis (lead); Funding acquisition (equal); Investigation (equal); Methodology (equal); Validation (lead); Visualization (lead); Writing – original draft (lead); Writing – review & editing (lead).

**Maki Shimizu**: Data curation (equal); Investigation (equal).

**Koji Sato**: Investigation (equal).

**Kisara Matsumoto**: Data curation (equal); Formal analysis (equal); Investigation (equal).

**Kazuki Ohnishi**: Investigation (equal); Resources (equal).

**Yongzhao Yao**: Funding acquisition (equal); Investigation (equal); Writing – review &

editing (equal).

**Ryo Hattori**: Funding acquisition (equal); Investigation (equal); Methodology (equal); Resources (equal); Writing – review & editing (equal).

## Data availability statement

The data that support the findings of this study are available within the article and its supplementary material.

Figure Captions

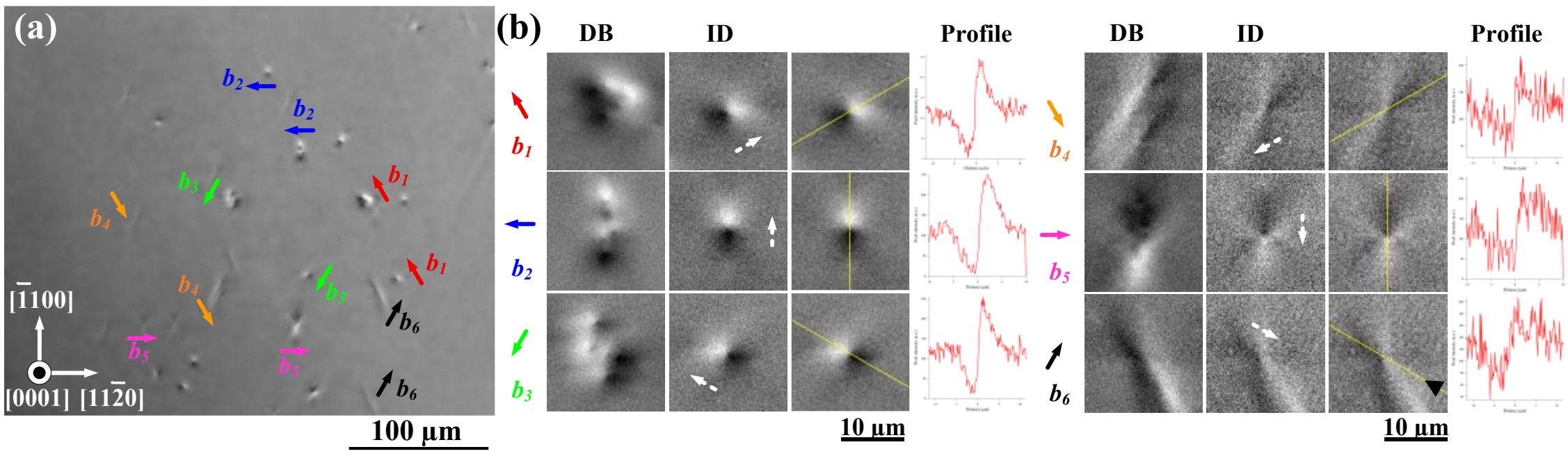

Fig. 1. (a) Phase-contrast microscopy image of a honeycomb defect formed in ammonothermal GaN. (b) For each of $\boldsymbol{b}_1$–$\boldsymbol{b}_6$, the panels from left to right show an enlarged image of the dislocation bundle, an isolated dislocation, the position used to extract the contrast profile, and the corresponding contrast profile.

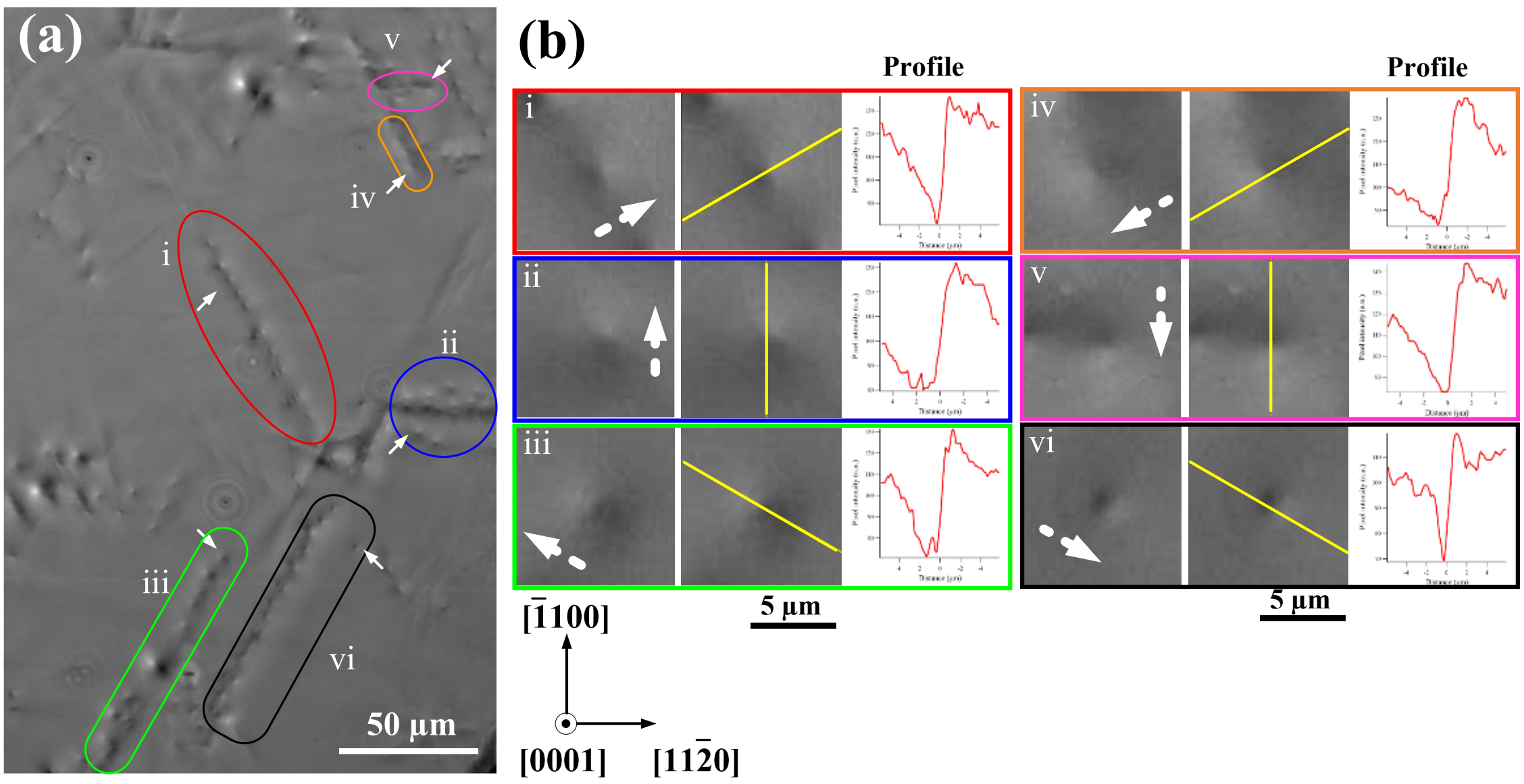

Fig. 2. (a) Phase-contrast microscopy image of HVPE-grown GaN. (b) For each contrast type (i–vi), the panels from left to right show an enlarged image of the dislocation, the position used to extract the contrast profile, and the corresponding contrast profile.

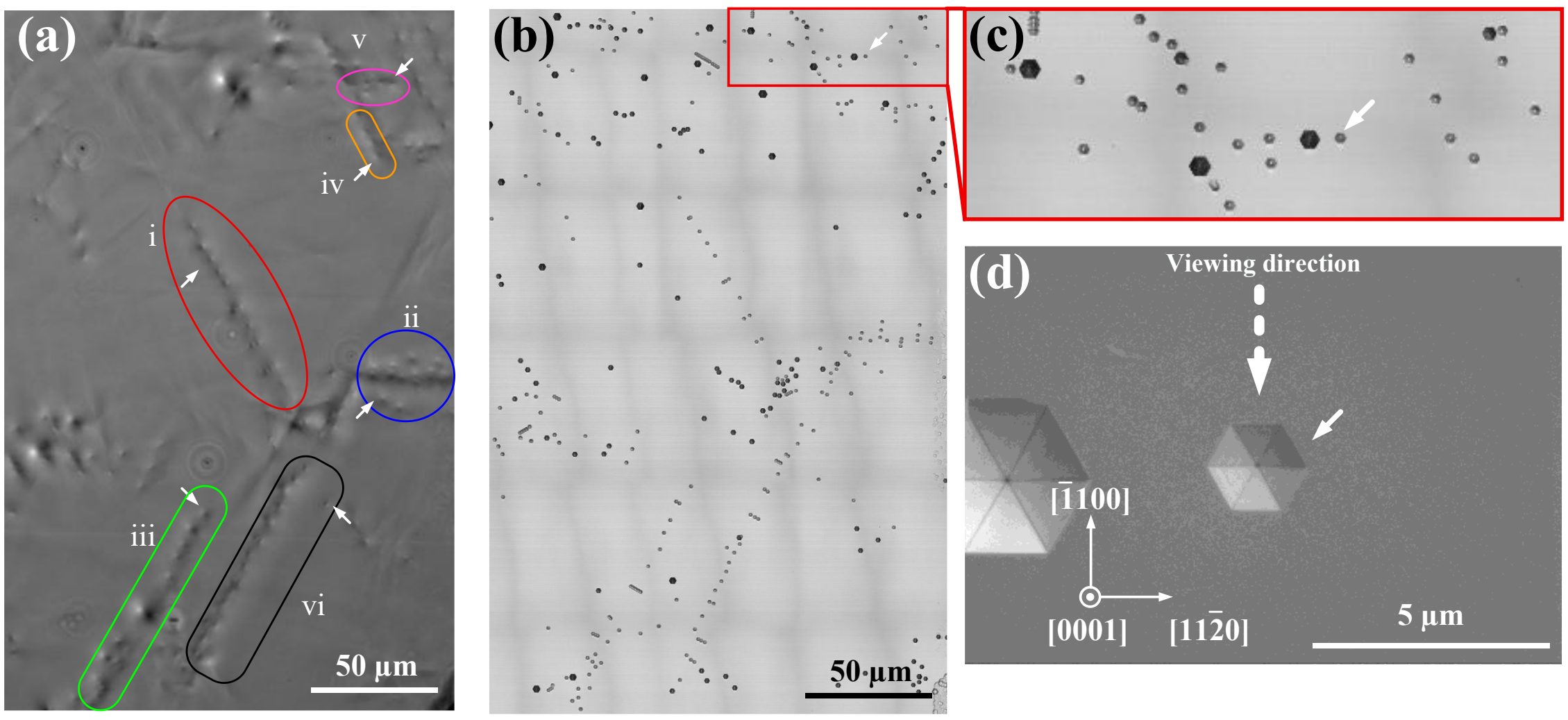

Fig. 3. (a) Phase-contrast microscopy image of HVPE-grown GaN [reproduced from Fig. 2(a)]. (b) Etch-pit image of the region shown in (a). (c) Enlarged image near region v in (b). (d) SEM image of the etch pit selected for TEM observation.

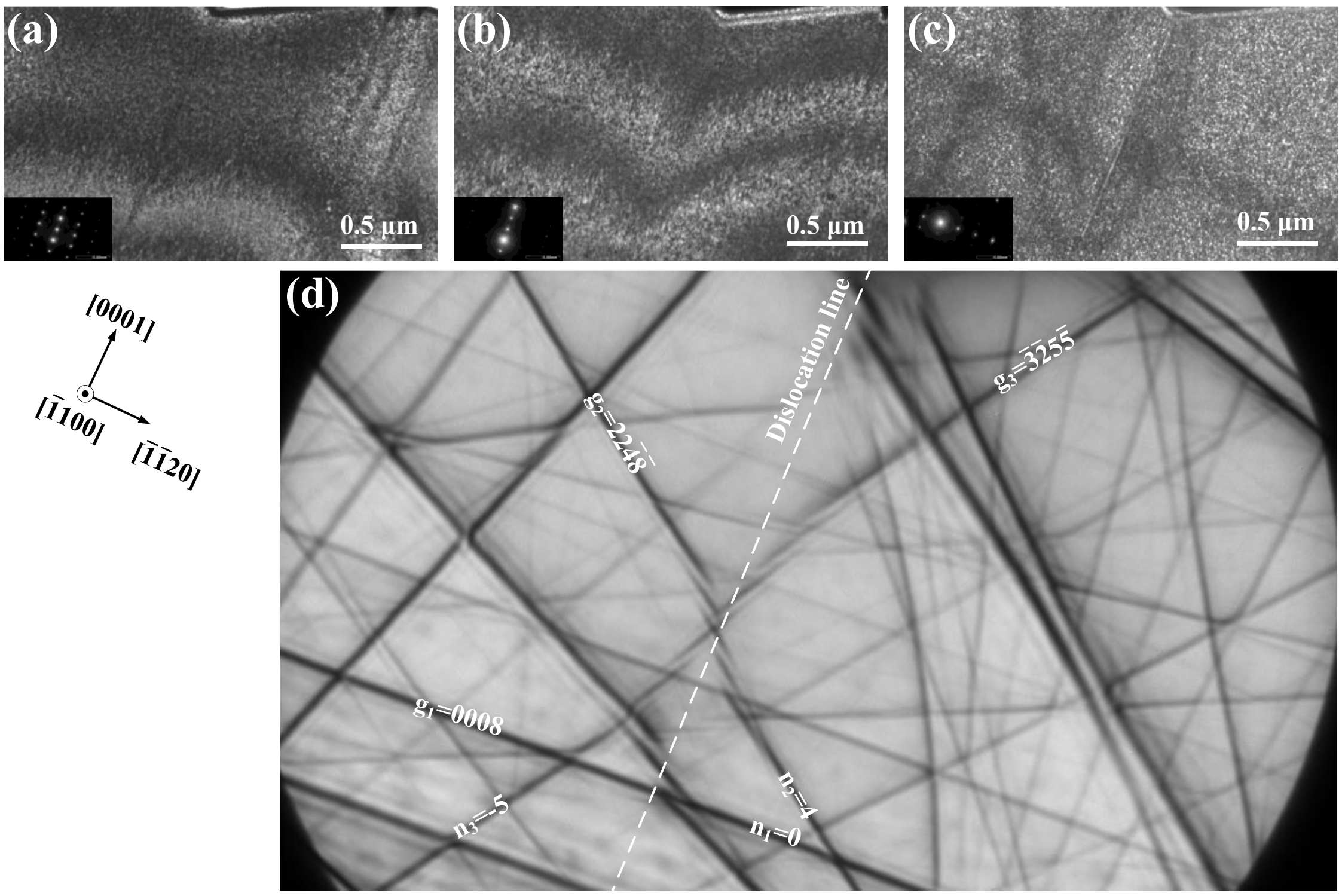

Fig. 4. (a) TEM image. (b), (c) Weak-beam dark-field images under ***g***–3***g*** conditions: (b) ***g*** = 0002 and (c) ***g*** = 11$\bar{2}$0. (d) LACBED pattern.

## Supplementary Material

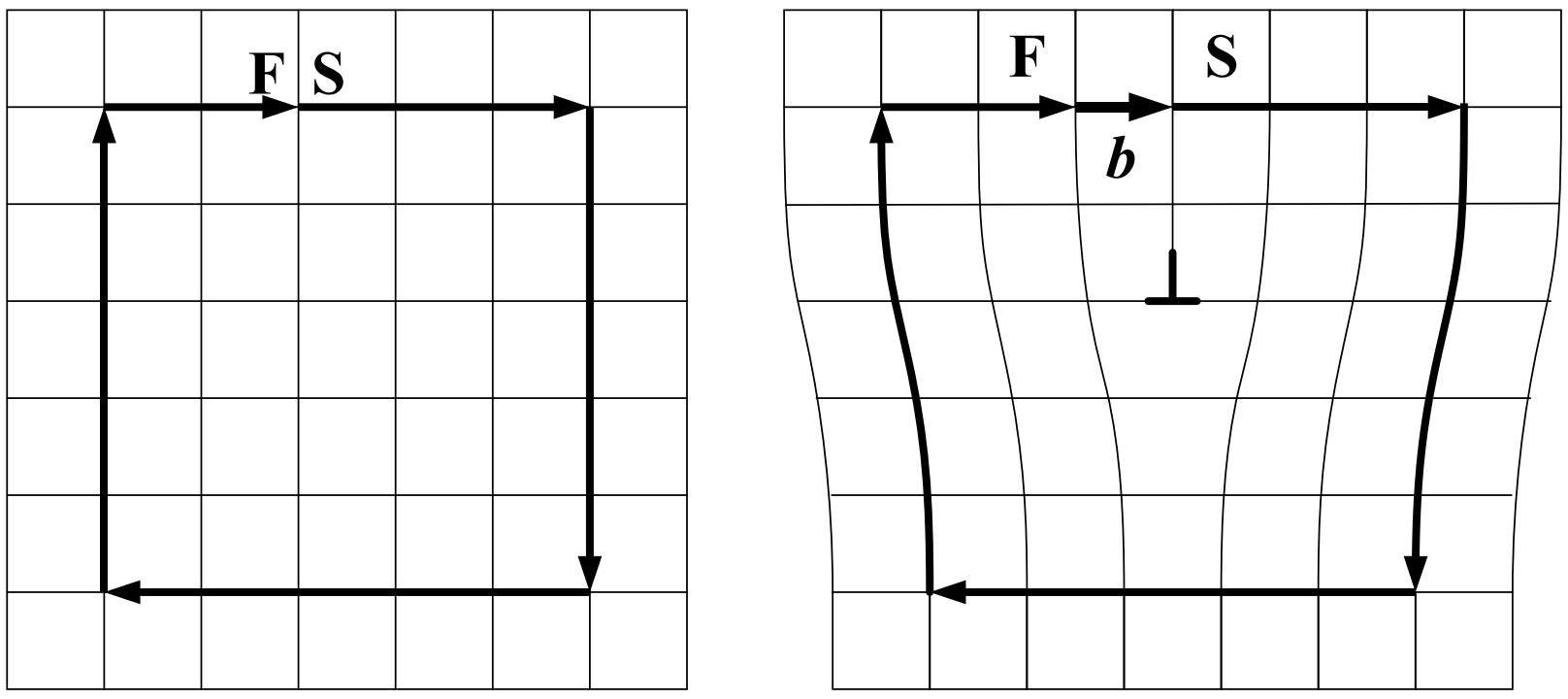

**Supplementary Fig. 1** Definition of the Burgers-vector convention used in this study. A closed right-handed Burgers circuit is constructed in the perfect reference crystal and mapped onto the dislocated crystal, where the finish-to-start closure vector defines the Burgers vector.

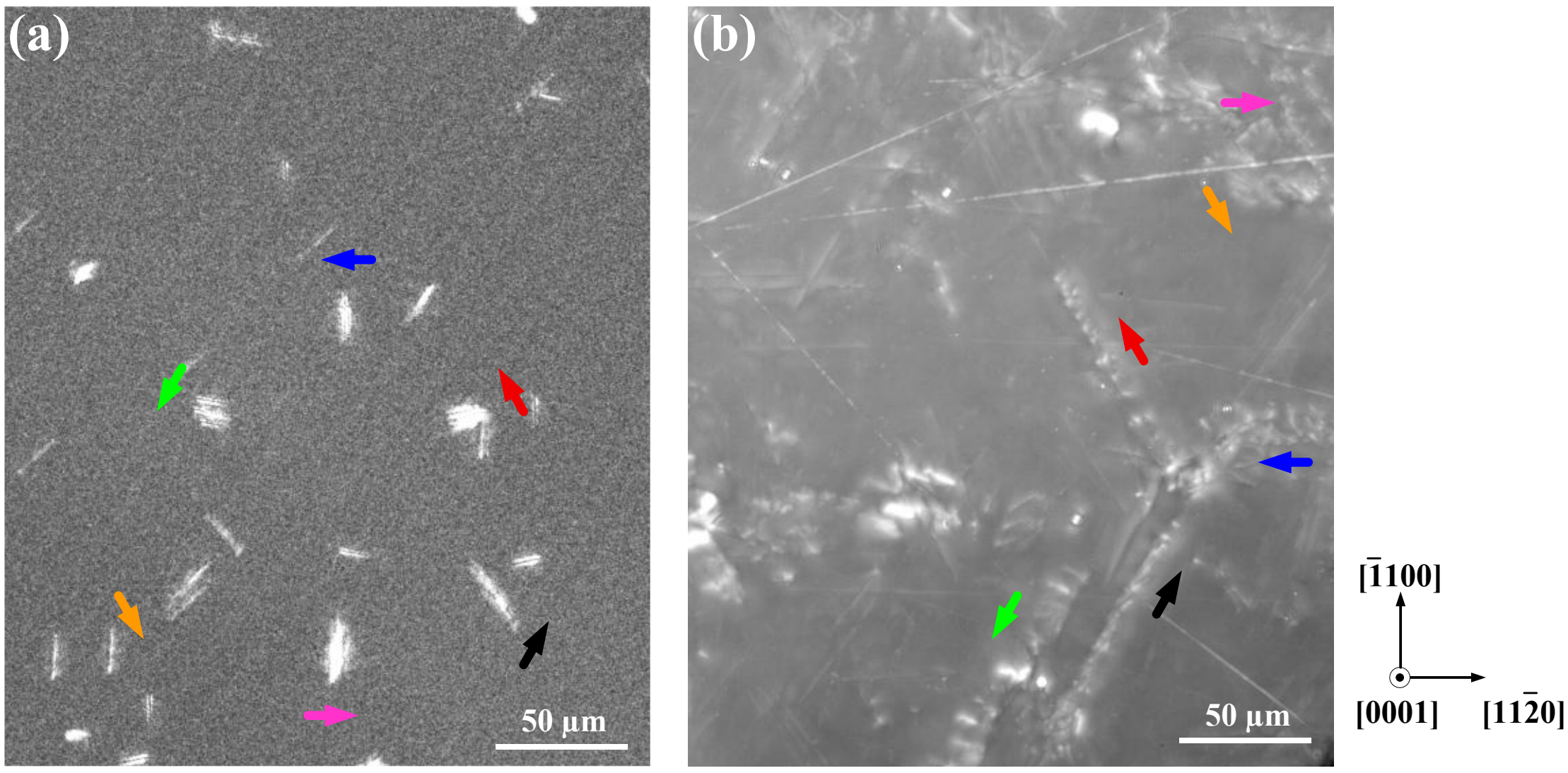

**Supplementary Fig. 2** (a) Projection image of dislocations in ammonothermal GaN reconstructed from polarized-light microscopy images. (b) Projection image of dislocations in HVPE-grown GaN reconstructed from phase-contrast microscopy images. Dislocations are aligned along six directions in ammonothermal GaN, whereas diverse morphologies are observed in HVPE-grown GaN.

**Order-of-magnitude estimate of the strain-induced phase shift**

Because experimentally determined photoelastic coefficients of GaN are unavailable, the following calculation is intended only as an order-of-magnitude estimate. A straight threading edge dislocation (TED) with its dislocation line parallel to [0001] and Burgers vector $\boldsymbol{b} = 1/3\,[11\bar{2}0]$ was assumed. The corresponding in-plane strain field was calculated using the anisotropic elastic solution for hexagonal GaN, and the refractive-index distribution was then obtained from the photoelastic relation described in the main text. The calculated refractive-index distribution in the *XY* plane was convolved with a Gaussian point-spread function (PSF) to account for the finite lateral resolution of the optical system. The standard deviation of the Gaussian PSF was taken as

$$\sigma_p = 0.21 \cdot \lambda/NA$$

which gives $\sigma_p$=0.170 μm for λ=405 nm and NA = 0.5.

To avoid the singularity of the continuum-elasticity solution at the dislocation core, a core cutoff radius of $r_0$=$\boldsymbol{b}$=0.3189 nm was introduced. Using the reported theoretical photoelastic coefficients $p_{11}$=0.031 and $p_{12}$=0.008 reported for GaN [Ref. S. Yu. Davydov and S. K. Tikhonov, Semiconductors **31**, 698 (1997).], the maximum magnitude of the PSF-averaged refractive-index change was estimated to be approximately

$$\left|\Delta \mathrm{n}_{eff}\right|_{max} \approx 1.1 \times 10^{-5}.$$

Using an effective axial information-acquisition depth of *l*=43 μm, the corresponding maximum phase shift,

$$\phi = 2\pi/\lambda \cdot \Delta \mathrm{n}_{eff} l,$$

was estimated to be

$$|\phi|_{max} \approx 7.6 \times 10^{-3} \text{ rad}.$$

Although the absolute magnitude depends on the photoelastic coefficients, which have not been experimentally established for GaN, this estimate indicates that the strain-induced phase modulation is much smaller than 1 rad under the present optical conditions. Therefore, phase wrapping or reversal of the phase-contrast polarity due to phase accumulation within the effective axial information-acquisition depth is unlikely within this model.

1 R. A. Oliver, Mater. Sci. Technol. **32**, 737 (2016).

2 X. Fan, J. Shi, Y. Chen, G. Miao, H. Jiang, and H. Song, Micromachines **15**, 1188 (2024).

3 H. Amano, Y. Baines, E. Beam, M. Borga, T. Bouchet, P. R. Chalker, *et al*., J. Phys. D: Appl. Phys. **51**, 163001 (2018).

4 M. Meneghini, R. A. Khadar, C. De Santi, L. Nela, I. Abid, M. Buffolo, *et al*., J. Appl. Phys. **130**, 181101 (2021).

5 M. Buffolo, A. Caria, F. Piva, N. Roccato, C. Casu, C. De Santi, *et al*., Phys. Status Solidi A **219**, 2100727 (2022).

6 S. Usami, T. Narita, Y. Kitano, T. Uesugi, T. Kondo, S. Yamaguchi, *et al*., Appl. Phys. Lett. **112**, 182104 (2018).

7 S. Usami, T. Narita, Y. Kitano, T. Uesugi, T. Kondo, S. Yamaguchi, *et al*., Jpn. J. Appl. Phys. **58**, SCCB24 (2019).

8 T. Narita, M. Kanechika, J. Kojima, H. Watanabe, T. Kondo, T. Uesugi, *et al*., Sci. Rep. **12**, 1458 (2022).

9 Y. Ishikawa, Y. Sugawara, D. Yokoe, K. Sato, Y. Yao, K. Watanabe, *et al*., J. Mater. Sci. **58**, 9221 (2023).

10 X. Ma and T. Sudarshan, J. Electron. Mater. **33**, 451 (2004).

11 T. Ouisse, D. Chaussende, and L. Auvray, J. Appl. Crystallogr. **43**, 122 (2010).

12 L. T. M. Hoa, T. Ouisse, D. Chaussende, M. Naamoun, and J. Achard, Cryst. Growth Des. **14**, 5761 (2014).

13 A. Tanaka, S. Inotsume, S. Harada, K. Hanada, Y. Honda, T. Ujihara, and H. Amano, Phys. Status Solidi B **257**, 1900553 (2020).

14 A. Kawata, K. Murayama, S. Sumitani, and S. Harada, Jpn. J. Appl. Phys. **60**, SBBD06 (2021).

15 S. Harada and K. Murayama, J. Appl. Crystallogr. **55**, 1029 (2022).

16 S. Harada, Y. Matsubara, and K. Murayama, Diamond Relat. Mater. **138**, 110192 (2023).

17 S. Harada, Y. Matsubara, S. Hayashi, M. Kawase, K. Seo, S. Mizutani, Y. Mizutani, S. Mizutani, and K. Murayama, AIP Adv. **15**, 025208 (2025).

18 K. Takahashi, Y. Matsubara, M. Kawase, K. Seo, S. Mizutani, Y. Mizutani, S. Mizutani, K. Murayama, and S. Harada, Acta Mater. **290**, 120923 (2025).

19 S. Harada and K. Murayama, J. Cryst. Growth **640**, 127982 (2025).

20 Y. Ishikawa, K. Matsumoto, K. Ohnishi, Y. Yao, arXiv:2606.27744

21 R. Hattori, O. Oku, R. Sugie, K. Murakami, and M. Kuzuhara, *Appl. Phys. Express* **11**, 075501 (2018).

22 R. Hattori, Y. Yao, and Y. Ishikawa, *Solid State Phenom.* **376**, 47 (2025).

23 Y. Ishikawa, D. Katsube, K. Sato, Y. Yao, and K. Sasaki, *APL Mater.* **14**, 021104 (2026).

24 Y. Ishikawa, R. Hattori , Y. Yao, D. Katsube, and K. Sato, Appl. Phys. Express 19, 041003 (2026).

25 L. Kirste, K. Grabianska, R. Kucharski, T. Sochacki, B. Lucznik, and M. Bockowski, Materials 14, 5472 (2021).

26 M. Tanaka, R. Saito, K. Ueno, and Y. Harada, J. Electron Microsc. 29, 408 (1980).

27 M. Tanaka, M. Terauchi, and T. Kaneyama, J. Electron Microsc. 40, 211 (1991).

28 D. Cherns and A. R. Preston, J. Electron Microsc. Tech. 13, 111 (1989). 20)

29 D. Cherns and J. P. Morniroli, Ultramicroscopy 53, 167 (1994).

30 J.-P. Morniroli, Large-Angle Convergent-Beam Electron Diffraction (LACBED): Applications to crystal defects, (Société Française des Microscopies, Paris, 2002).

31 L. Kirste, T. N. T. Caliste, J. L. Weyher, J. Smalc-Koziorowska, M. A. Zajac, R. Kucharski, T. Sochacki, K. Grabianska, M. Iwinska, C. Detlefs, A. N. Danilewsky, M.

Bockowski, and J. Baruchel, Materials 15, 6996 (2022).

32 N. Kokubo, Y. Tsunooka, F. Fujie, J. Ohara, K. Hara, S. Onda, H. Yamada, M. Shimizu, S. Harada, M. Tagawa, and T. Ujihara, Appl. Phys. Express 11, 061002(2018).

33 N. Kokubo, Y. Tsunooka, F. Fujie, J. Ohara, S. Onda, H. Yamada, M. Shimizu, S. Harada, M. Tagawa, and T. Ujihara, Appl. Phys. Express 11, 111001(2018).

34 N. Kokubo, Y. Tsunooka, F. Fujie, J. Ohara, S. Onda, H. Yamada, M. Shimizu, S. Harada, M. Tagawa, and T. Ujihara, Jpn. J. Appl. Phys. 58, SCCB06 (2019).

35 Nobuhiko Kokubo, Yosuke Tsunooka, Sho Inotsume, Fumihiro Fujie, Shoichi Onda, Hisashi Yamada, Mitsuaki Shimizu, Shunta Harada, Miho Tagawa, and Toru Ujihara, Jpn. J. Appl. Phys. 60, SAAD03 (2021).

36 P. M. Anderson, J. P. Hirth, and J. Lothe, *Theory of Dislocations*, 3rd ed. (Cambridge University Press, Cambridge, 2017), p. 68.

37 N. Gmeinwieser and U. T. Schwarz, Phys. Rev. B 75, 245213 (2007).